\documentclass{article}
\usepackage{authblk}
\usepackage{graphicx} 
\usepackage[left=2.4cm,right=2.4cm,top=2.4cm,bottom=2.4cm] {geometry}
\usepackage[utf8]{inputenc} 
\usepackage[T1]{fontenc}    
\usepackage{hyperref}       
\usepackage{url}            
\usepackage{booktabs}       
\usepackage{amsfonts}       
\usepackage{nicefrac}       
\usepackage{microtype}      
\usepackage{cite}
\usepackage[version=4]{mhchem}
\usepackage{doi}
\usepackage{geometry}
\usepackage{tabularx}
\usepackage{caption}
\usepackage{multicol}
\usepackage{booktabs} 
\title{Challenges and Opportunities in Implementing Negative Differential Resistance Mode Reconfigurable Field Effect Transistors}
\author[1]{Lephe S}
\author[1]{Gifrin Fredik Raj S}
\author[2]{Abijesh Euphrine A}
\author[1]{Janaki S}
\author[3]{Jamina C}
\author[1\thanks{Corresponding author.}]{Arun Jose L}
\affil[1]{Department of Physics, St. Xavier’s College, Affiliated to Manonmaniam Sundaranar University, Palayamkottai, Tamil Nadu 627002, India}
\affil[2]{Deen Dayal Upadhyay Centre, Loyola College, Affiliated to University of Madras, Chennai, Tamil Nadu 600034, India}
\affil[3]{Sethu Lakshmi Bhai Government Higher Secondary School, Nagercoil, Tamil Nadu 629002, India}

\date{\today}

\begin{document}

\maketitle
\begin{abstract}
Desirably, the world relies on the devices being compact, as they could drive to the increased functionality of integrated circuits at the provided footstep, that is becoming more reliable. To reduce the scalability over the devices, approach has been outlined utilizing the NDR mode reconfigurable functionality over the transistors. Being an individual device efficient in exhibiting different task with the different configurations in the same physical circuitry. On the view of reconfigurable transistors, possibly authorize the reconfiguration from a p-type to n-type channel transistor has been expelled as an emerging application such as static memory cells, fast switching logic circuits as well as energy efficient computational multi valued logic. This article emphasizes NDR mode RFET along with its classification, followed by enhancing the RFET technology concepts and RFET’s future potential has been discussed briefing with the growing applications like hardware security as well as neuro-inspired computing.
\end{abstract}

\section*{Keywords}
reconfigurability; NDR mode RFET; neuromorphic computing; hardware security

\section{INTRODUCTION}
Owing to the increased power consumption in the electronic circuits, a reconfigurable switching element specifically, negative differential resistance(NDR) paves an  approach for conceiving alternative current (a.c.) electricity, that would be feasible to generate it.  \cite{baldauf_stress-dependent_2015-1,merten_quality_2022,trommer_elementary_2013}  Transistors alongside with the computer technologies amidst communication and information could be performed adopting lesser power and space  \cite{krinke_exploring_2021} . Reconfigurable field effect transistors (RFET) an electronic device  using NDR, exhibits basic switching functions like n-type and p-type FET operations suchness gate tunable performance using external terminal voltages, as such reconfigurable transistors are basically two devices comprised of one  \cite{pregl_printable_2016, sessi_junction_2018} . The functional reconfiguration property rely on the electrostatic doping, i.e. electron potential-based production of mobile electrons rather than the chemical doping, whose property alters due to the impurities being added \cite{kang_2d_2020, mikolajick_rfetreconfigurable_2017} .

\begin{figure}
    \centering
    \includegraphics[width=13cm,height=10cm]{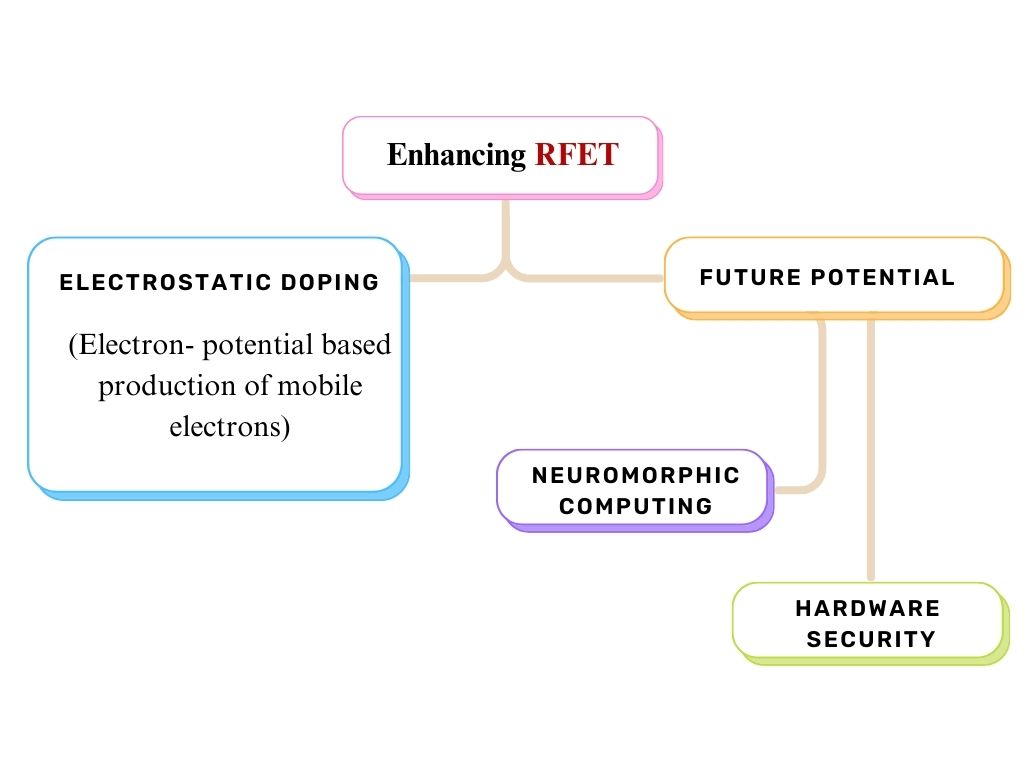}
    \caption{Flow chart for enhancing RFET}
    \label{fig:1}
\end{figure}

The efficiency of the circuit has been ascended ahead of the limitations being encountered, the ability of transistors to reconfigure has witnessed computational enhancement over the field \cite{sessi_junction_2018,larentis_reconfigurable_2017,trommer_functionality-enhanced_2015} . The reconfigurability of the transistors devices has been disciplined by two gate electrodes: a). control gate electrode \cite{de_marchi_polarity_2012} , particularly to modulate drain current thermionically through transistor to turn the device on and off, b). polarity gate electrode \cite{zhang_polarity-controllable_2014} , which sets the charge carrier type in the channel to swap within p and n type operation in transistors. The switching ability i.e. reconfigurability elevates the imperative need of the basic building blocks with the rigorous polarity at the destined location in the systems \cite{darbandy_high-performance_2016} . Unalike the traditional complementary metal oxide semiconductor (CMOS) based transistor, reconfigurable transistors overcomes the elementary circumspection, which could increase the utility of the electronic systems rather than to increase the soaring \cite{alasad_logic_2017, chen_reconfigurable_2021,heinzig_dually_2013,navarro_reconfigurable_2017} .

The article commences with the concepts of RFET along with the basic requirements of RFET fabrications in the section 2. In section 3, deals with the history of RFET’s and its advancements. In section 4, RFET and its classification along with its Schottky and non-Schottky approaches has been discussed. Followed by section 5, negative differential resistance has been briefed. In the section 6, deals with enabling as well as enhancing the technology over RFET has been discussed. Alongside, section 7, incorporates some of the add on features like polarity control with the non-volatile memory has been enlisted in the section 8. Section 9 focuses on the potential of RFET and its future applications such as neuro-inspired computing as well as hardware security.

This article could elaborately showcases the NDR assisted RFET devices and its concepts of achieving NDR along with the advancements required to enhance the RFET technology along with its future applications such as hardware security as well as neuro-inspired computing over another article being published on the front end. 

\section{CONCEPTS OF RFET}
To view the electrical behaviour of RFET, unipolar device characteristics of various RFET concepts has been summarized. Apart from the different RFET concepts, new concepts has been proclaimed, as carbon nanotubes, graphene and silicon nanowires has been implemented for the approach \cite{darbandy_high-performance_2016} . In Schottky contacts, low doped semiconductor materials or intrinsic semiconductors are frequently passed down for the metal/semiconductor interfaces. And as in the new approach over RFET, Schottky contacts as well as the gate electrodes in the active region will be efficiently controlled, active regions over the electrode is electrostatically tenacious \cite{pregl_printable_2016}.
Considering, the efficiency over the gate electrodes in RFET concept aims attention at the selection of polarity gates at the contacts as well as the complete control over the amount of charge injected on the channel. Weak gate coupling occurs as a reason of buried gates together with common gates, conjointly, high operational voltage is required for the RFET devices \cite{bockle_gate-tunable_2021} . Integrating the device to two outer top gates instead of singular back gate, which functions simultaneously at the same potential as same as the back gate functions, though it accelerates the limited scalability over the device structure that authorizes an individual control over the different functionalities over a single chip, yet they provide adequate gate coupling over the entire region by the gates placed at the outer for both control as well as program gates. Plenty of nanowire channels has been deformed in order to inflate on current ratio \cite{sistani_nanometer-scale_2021} .
In order to incorporate electron and hole injection over the contacts, a Schottky barrier is introduced. To enable the reconfigurability of the devices, two Schottky junctions has been introduced to the two sides of the nanowire heterostructure. Along with the Schottky junctions, improving the gate control over the active region enhances the nanowire heterostructure. The control as well as the program  gates are placed at the two scottky junctions as they are programmed to select either p or n type configuration\cite{heinzig_reconfigurable_2017}.

\begin{figure}
    \centering
    \includegraphics[width=6cm,height=6cm]{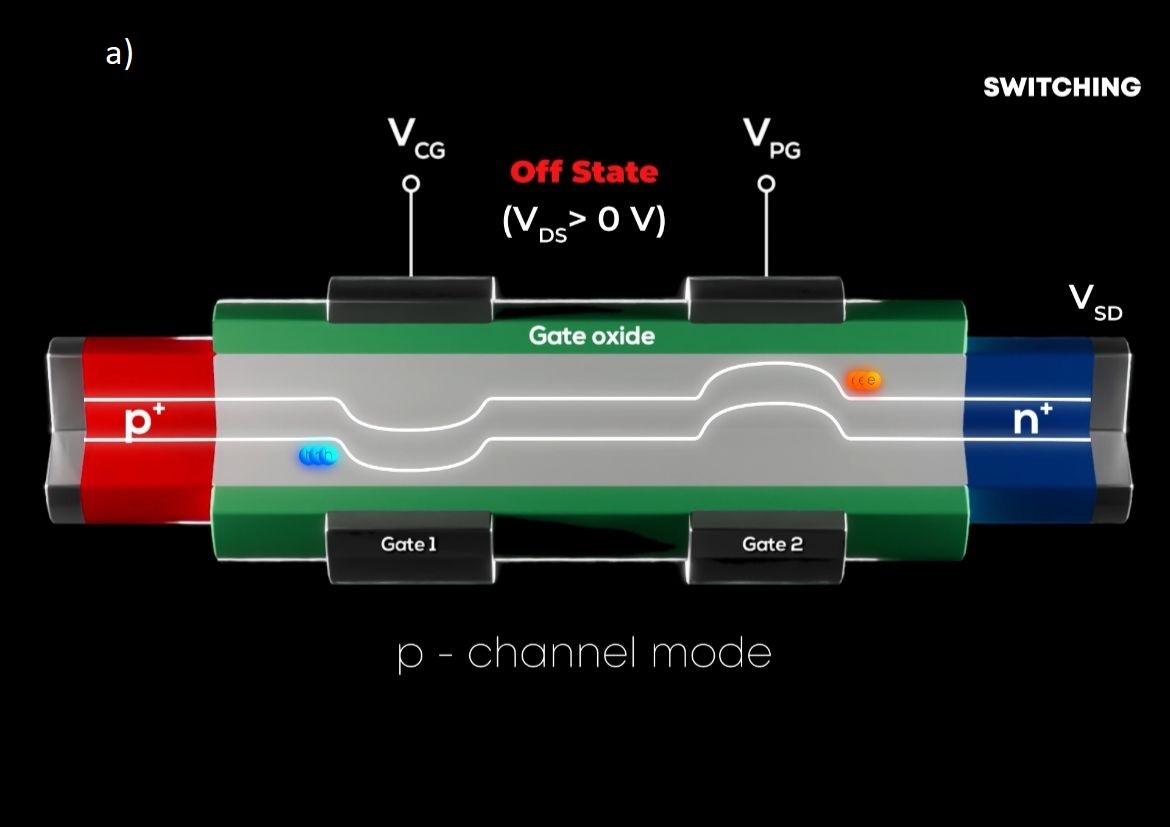}
    \includegraphics[width=6cm,height=6cm]{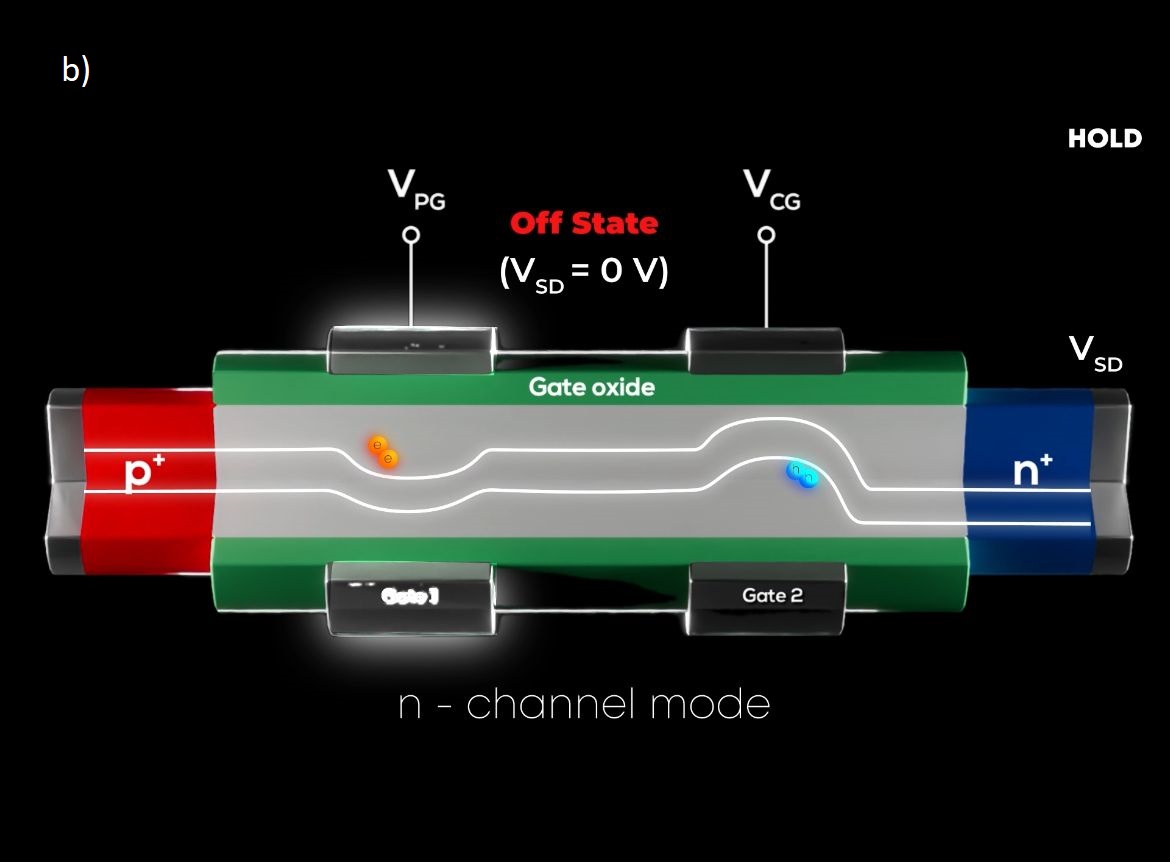}
    \includegraphics[width=6cm,height=6cm]{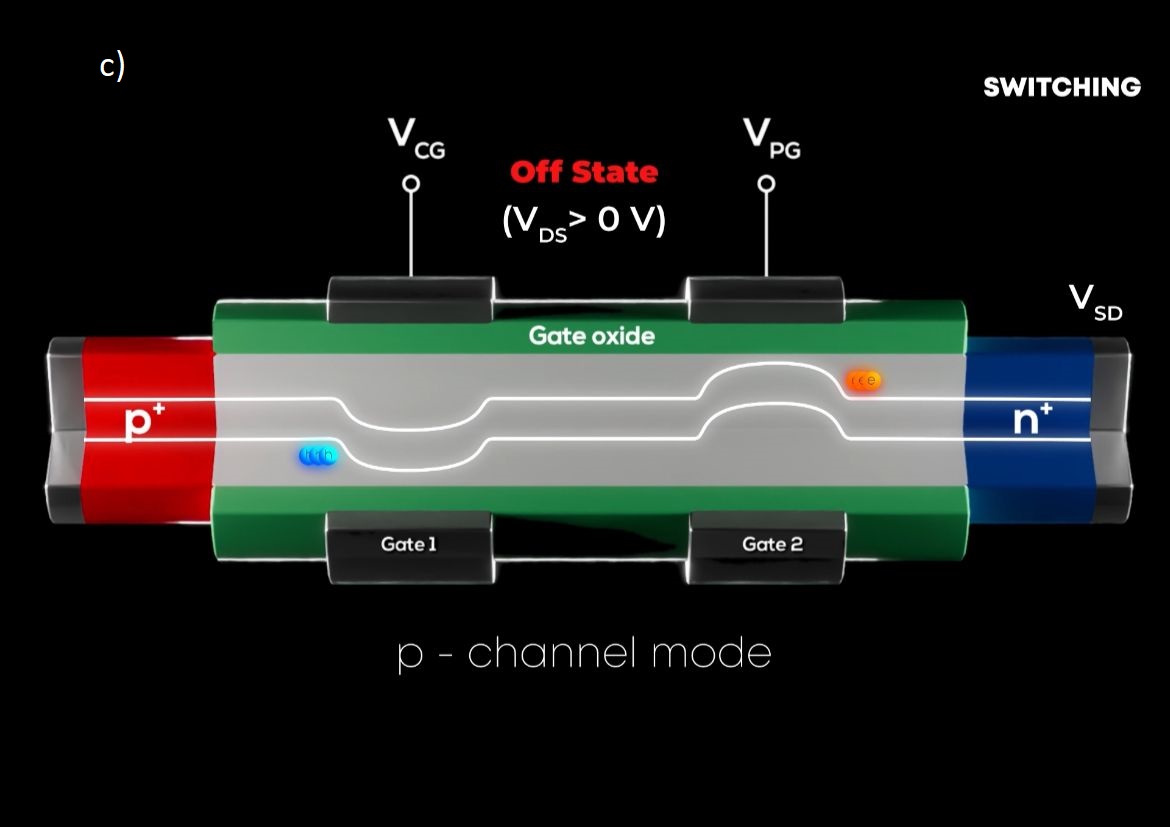}
    \includegraphics[width=6cm,height=6cm]{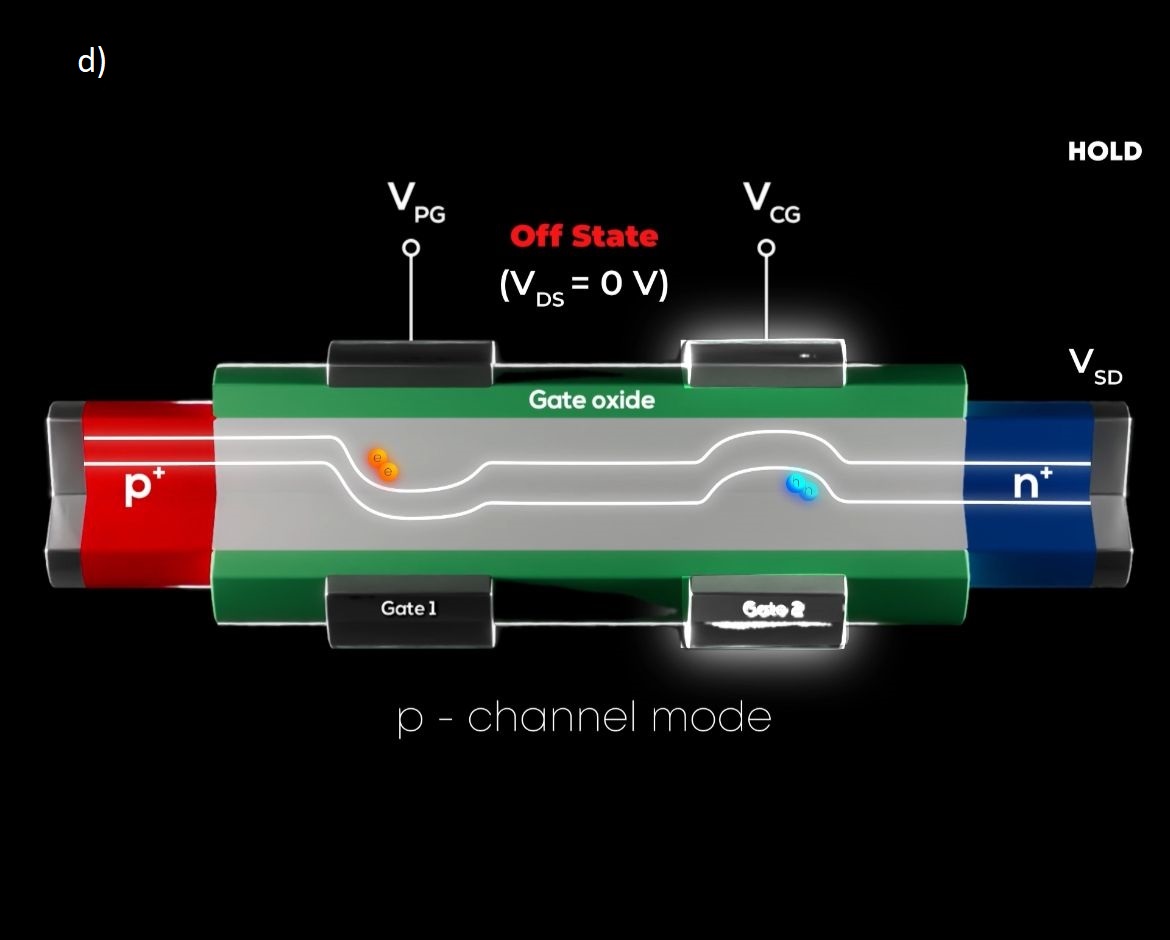}
    \caption{The reconfigurable silicon nanowire FET with the independent junction gating, two independent gates have been pinpointed on top of the source and drain junctions. a) showcases the off state switching in n-channel mode at $V_{SD}$ = 0 V, b) represnts the off state hold in n-channel mode at $V_{SD}$ = 0 V, c) represents the off state switching in p-channel mode at $V_{DS}$ > 0 V, d) shows the off state hold in p-channel mode at $V_{DS}$ = 0 V.}
    \label{fig:2}
\end{figure}

In off state, by the dint of Schottky junctions, charge carriers have been inoculated within the top gate and gets trapped over channel, as far as the control gate remains unblocked. As long as the injected charge carriers being trapped within the gates, provokes the device to degrade \cite{mongillo_multifunctional_2012} . Leverage on using independent gating control over the Schottky contacts is that charge carriers can be blocked at the Schottky interface even before reaching the active regions, further the polarity of the device remains unchanged \cite{baldauf_tuning_2017} . 
Considering the RFET concepts, independent gating could possibly be relative to large fraction of channel ungated, impact over the ungated regions has been studied for measurements. RFET with the common back gate electrodes preceding the fabrication of top gates are compared with the measurements with the uses of two top gates as shown in figure 2. \cite{roy_insightful_2022} . As ungated regions do not hinder the transport properties, when compared with saturated on and off currents. When charge carrier density distribution has been computed over, the charges travels from the source to the drain, as the charge track has been erected to follow a sand glass shaped trajectory \cite{kundu_design_2022-1} . Since, the charge carriers has been injected into the gated regions, nonetheless, it penetrates into the ungated regions, similarly they bundle near the centre of the nanowire, holds for both p and n operations.
For the reason that, inverse subthreshold slope limit has been smoother to accomplish with the concept of RFET, besides one side of the band is bound to the junction, establishing the shape of the parabolic  barrier as complicated one \cite{zhao_nonvolatile_2021}. Without altering the barrier width at the control gate at the Schottky contacts, the control gate over the Schottky junction has been decoupled. Schottky FET’s concept has been briefly described in detail in the section 4 \cite{tung_recent_2001} .

Meanwhile, when RFET remained turned off, thermionic emission takes place, on the contrary, changes in the tunnelling transmission takes place until the flat band shape is attained \cite{baldauf_scaling_2019}. The barrier over the channel is controlled by its full width.

Conceptually, RFET’s principle focuses on one potential barrier to distinct the two states individually similar to the case of conventional Metal Oxide semiconductor field effect transistor (MOSFET) exhibiting large scalability than devices consisting of two or few more potential barrier in series \cite{reuter_mosfets_2020} . Also, the scalability over the device is determined by the tunnelling transmissibility of the potential barrier being very thin.

\section{HISTORY OF RFET}
RFET originates in order to overcome the undesired high off-currents as in the Schottky barrier thin film transistors and came up with the solution of accomplishing a device with two gates, one, the control gate (CG) \cite{khan_controlled_2021} , is present to gate the source sided-Schottky junction, alongside the gate and the source sided Schottky junction are wrapped with thick oxides. Another stands as program gate (PG), which serves as a sub-gate, placed in order to complete the transistor, also assists in excluded carrier injection from the drain side in off-state as that could lowered by the order of magnitude, accounts on high on/off current ratio in the order $10^6$  \cite{park_analysis_2022} .  Following the TFT with polycrystalline silicon, the later equipped with separation by implanted oxygen (SIMOX) wafers with monocrystalline silicon, alongside bringing forth upgraded architecture performing maximum on/off current ratio of the order $10^9$ and an accountable sub threshold swing at 60mV/dec, however still in desperate need of PG relevantly higher than the operating voltage \cite{trommer_elementary_2013,mongillo_multifunctional_2012,li_impact_2021} . RFET concepts has been experimentally demonstrated, in particular on a target application it was perceived, gained attention in order to reduce the scalability of the device sizes in accordance with the Moore’s law \cite{mikolajick_silicon_2013} . On the trend of minimizing the scalability of the device, reliable and substantial MOSFET has been debated long over the nanoscale dimensions. Introducing dopant over the ultra-scaled technology nodes, incorporation of chemical impurities added up an impact over the nanoscale channels with the dopant deactivation technique \cite{reuter_mosfets_2020,sistani_bias-switchable_2021,zheng_proposal_2021} . Later on, the introduction of electrostatic doping over the devices delivers promising ultra-sharp junctions with well-versed carrier concentration as well as minimal defects over the device.

Subsequentially, RFET’s has been passed down with the carbon nanotubes \cite{yu_analysis_2022} , which acts as the channel material incorporated with control gate at the top and a program gate enclosing the whole channel from the backside of the channel, consciously PG voltage is applied synchronously at the Schottky junctions. From the figure 2.a. the logic gates NAND as well as NOR can be easily reconfigured with the replacement of drain with the gate and vice-versa through the voltage control over the devices, alongside multiple XOR gates can be built with the least number of transistors.

In consideration of the voltage signs being applied, the polarity is set to bend over the bands in the conducting channel region, i.e., positive voltage allows the injection of electrons to the entire channel as well as negative voltage allows the holes to circulate within the entire channel \cite{li_impact_2021} . Additionally, the control gate over the channel induces thermionic energy barrier which could access and modulate the channel which could switch the transistor on/off. Since the charge carrier are injected into the channel via the Schottky junctions as the control gate operates, particularly at room temperature, the device attains a slope of about 60mV/dec, though the structure could be repeated for ambipolar operational devices for the selective carrier control, along way the back gate is employed for controlling the control gate injection, based on the voltage applied, so is the type of carriers injected \cite{baldauf_scaling_2019,roemer_uniform_2021} . To steer the channel, the buried control gates exhibit an ambipolar behavior, but in the case of program gates, they are placed at the top of the gate to block and remove the unwanted carrier injection within the whole channel. Conceptually, RFET’s has been improved by exploring heterostructure ($NiSi_2$/ Si/$NiSi_2$) in conjunction with the silicide contacts, where no intermetallic phases between Al-Si interface has been spotted, alongside the gate assigned precisely at the drain barrier to chunk the unwanted carrier type, as this exhibits lower source drain leakage as well as high sub threshold slope \cite{khan_controlled_2021,tang_solid-state_2014} . Following the concepts till date, Schottky-based RFET still confides a merging of both programming mechanisms and either one of them. Improvising the mechanisms, the focus turned towards the 1-D nanostructures, on the whole focused on the Si nanowires \cite{rai_physical_2018} . Later on, the device was refined with perfect p and n type current indispensable for the efficient usage of the reconfigurability in the CMOS like circuits, achieved by occupying oxidation induced mechanical stress into the channel which could enable operation with only two distinctive potentials, also the operation has been successfully carried out for the first reconfigurable complementary inverter circuits \cite{larentis_reconfigurable_2017} .
\begin{figure}
    \centering
    \includegraphics[width=6cm,height=6cm]{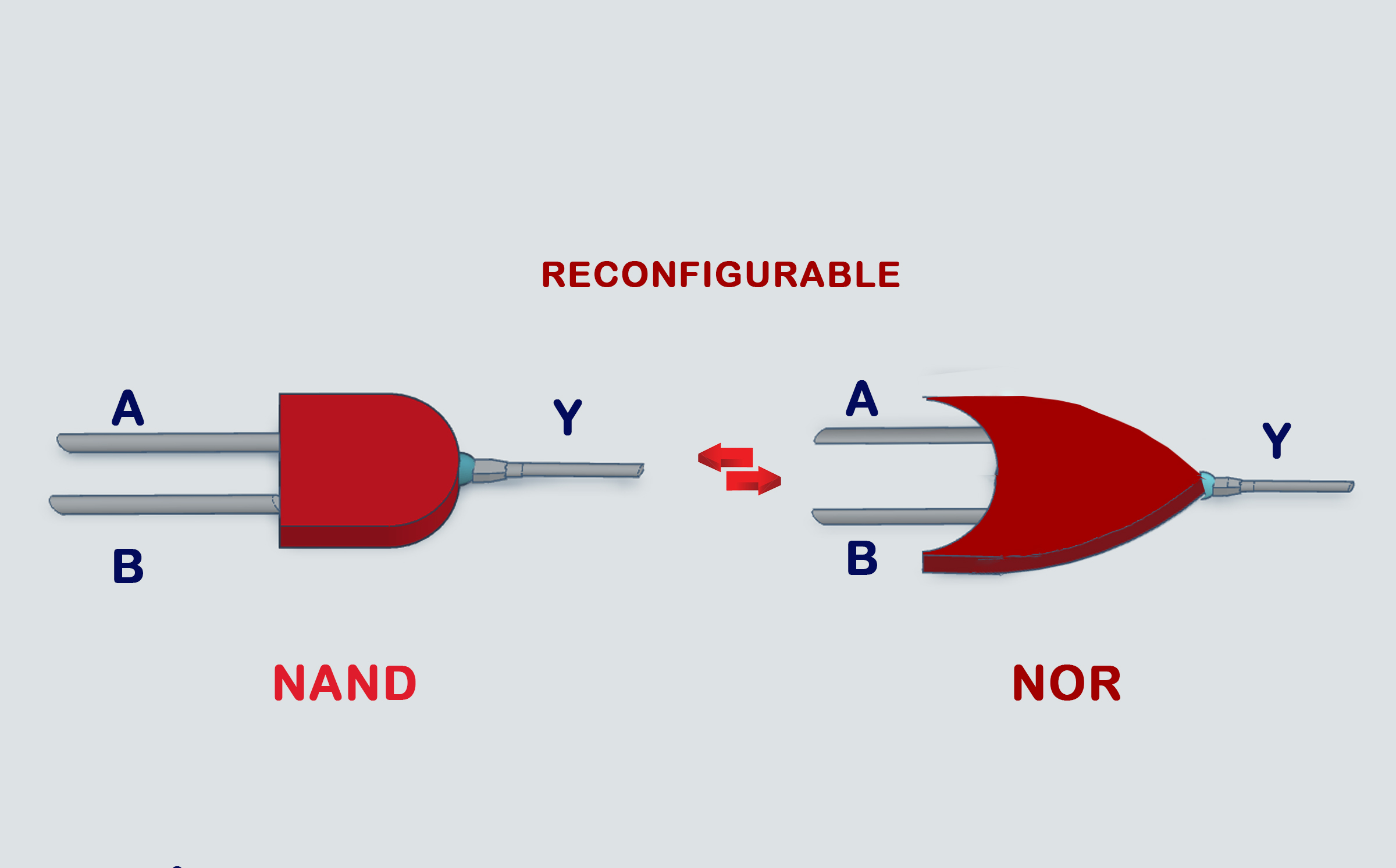}
    \includegraphics[width=6cm,height=6cm]{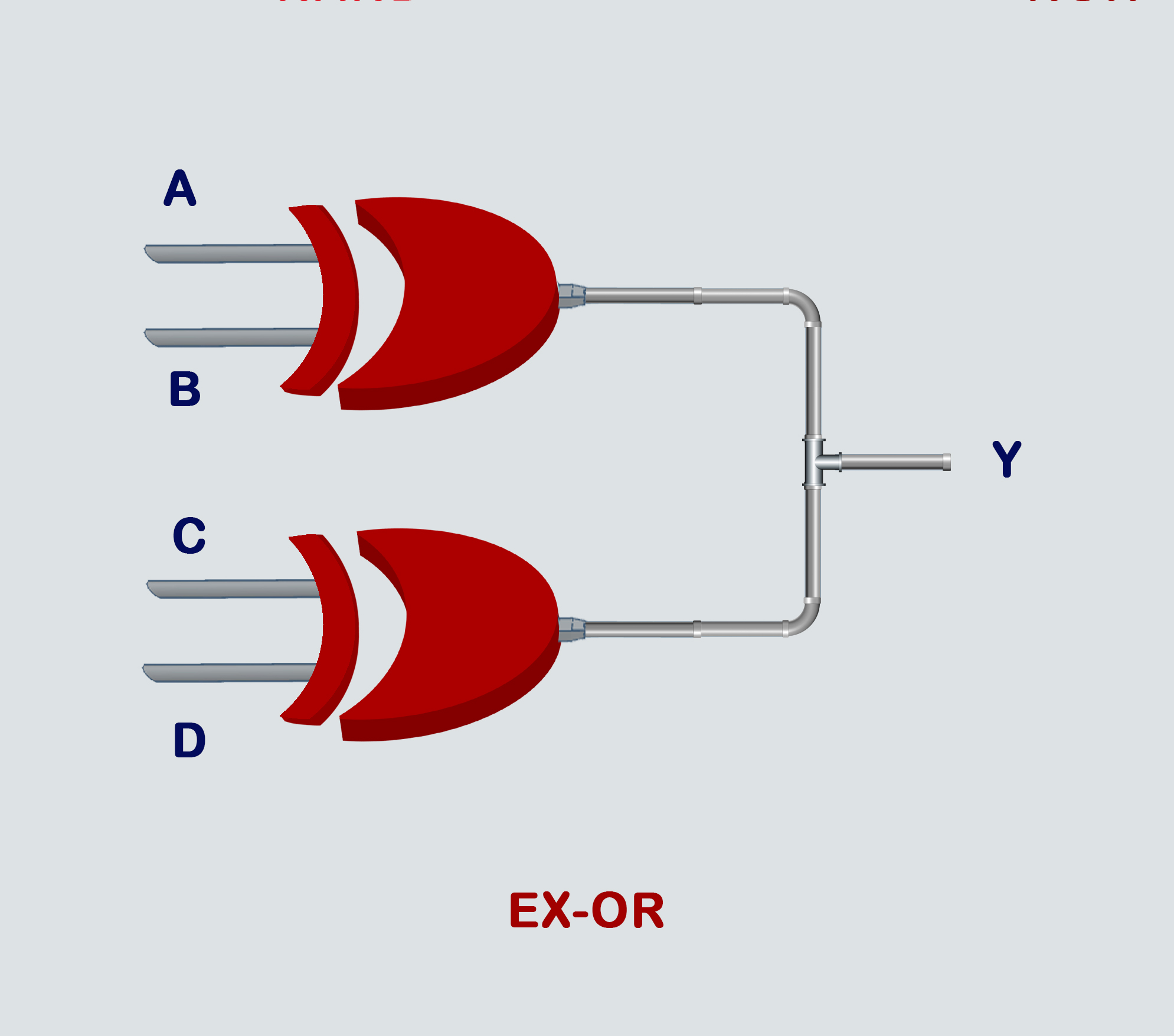}
    \caption{Represents the explicit and implicit reconfigurability on the logic gate level. a) Shows the reconfigurable NAND and NOR gate. b) Functional XOR gate.}
    \label{fig:3}
\end{figure}
Later on, three gated device with two outer gates protruding the Schottky junctions enhances the concept of back gate programming, for the reason of device integration as well as to reduce the programming gate voltages \cite{trommer_reconfigurable_2016} . Sophisticatedly, development over the device demonstrators leads to the first logic gates in conjunction with the RFET concepts. So, as to achieve more functionality per computational unit, two major archetypes can be acclaimed for the logic optimization level as a). implicit reconfigurability, utilized in logic gates, whereas a combination of inputs results in the truth table holding highly expressive capability b). explicit reconfigurability, which could amend the functionality over the external signal. For instance, NAND gate has been reprogrammed as NOR gate functionalities, which has been built from the RFET concepts \cite{galderisi_reconfigurable_2022,baldauf_vertically_2018} . As an instance, three gated RFET’s, lesser number of transistors has been used for XOR than in the CMOS, which is represented in figure 2. Complimentary circuit level features has been authenticated with some added benefit over the CMOS counterparts like control over the threshold voltage, intrinsic XOR, wired AND capabilities, and the suppression of charge sharing effects in dynamic logic gates \cite{alasad_logic_2017,heinzig_dually_2013,weber_reconfigurable_2014} . The add on functionality over the RFET concept happens over the circuit level than the device level. Along with the increased amount of advancements over the circuit level, in terms of development over the single device started to assorted in terms of material and transport like graphene, as well as black phosphorous and transition metal dichalcogenides has been focused on list over years \cite{kundu_design_2022}. Circuit fabrication on small scale demonstrations with the assistance of base material catches up with the silicon technology \cite{weber_silicon-nanowire_2006} . 
 
Circuitry design based on the RFET are intended in the advancement of accurate models and the support over the electronic design automation \cite{sessi_silicon_2020} . Behavior of the classical MOSFET’s has been developed reflecting the concepts of the device. Standardized cell materials has to be treated expressively, for the credit capability of the RFET technology \cite{zhang_novel_2022} . A pre circuit analysis is mandatory for the compassionate on adequacy of the emerging technology.  Various works has been reported for the reconfigurable nanotechnologies, as it puts forward transistor level models to carry out analysis over the simple circuits, as this models helps in easy accessible of transistor behavior at circuitry level \cite{khan_future_2020} .

\begin{figure}
    \centering
    \includegraphics[width=11cm,height=9cm]{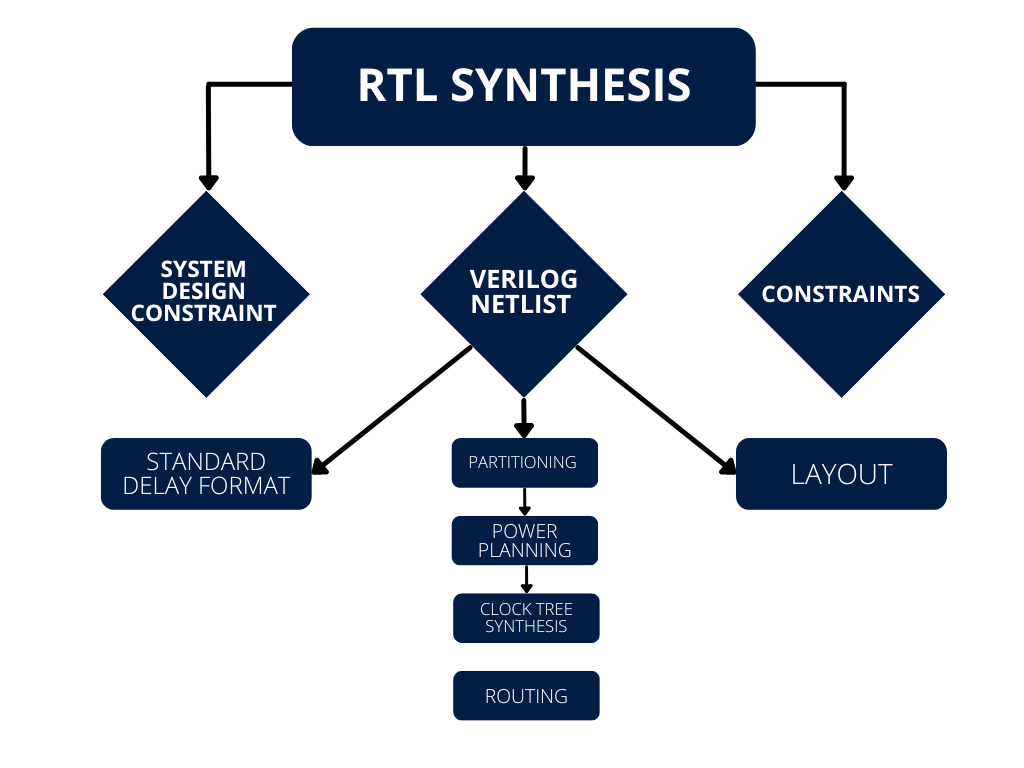}
    \caption{the top-down electronic structure automation (EDA) flows are imperative throughout the advancement of RFET technology, also they require exploring technology and can withhold unique properties} .
    \label{fig:4}
\end{figure}

The automation of electronic device design automation constitutes of two diversifications as a). logic synthesis as well as b). physical synthesis. In recent years, various approached has been made to enhance the RFET characteristics with automated circuit design along with the electronic design automation (EDA) flow and various approaches has been enlisted in the figure 3, alongside netlist over the EDA flow includes the ports, pins and blocks list also each of this has been enlisted with the individual identifier with it \cite{edwards_real_2021} . Currently, software has been developed to manage, verify and create system design constraints \cite{rai_technology_2018} . 

Considering the logic synthesis, RFET’s comprised of self-dual logic gates has been an outstanding choice for the cell standards are efficiently employed, also this stands for the technology mapping stage as well as the physical synthesis \cite{rai_technology_2018} . To yield the full potential over the features, new data has been required, alongside the logic synthesis accounting to the self-duality during the logical optimization results in the better area for RFET based circuitry devices \cite{neto_lsoracle_2019} . To achieve the touchstone level analysis, top-down EDA flow has been opted, also required to enable the standard cell-based application specific integrated circuits (ASIC) flow for the upcoming technological advancements.

Some of the works of complete physical synthesis flow for the emerging reconfigurable nanotechnologies has been reported accordingly, as the demonstration consisting of 22nm nanowire which consists of silicon model utilized for the characterization of common logic gates like AND, OR, MUX, XOR as they can also be used in the standard cells as in typical EDA flow \cite{fan_enabling_2022} . Followingly, many concepts have been reported along with the parallel RFET technology proposed with the standard cell synthesis over the demonstration. 
\section{RFET CONCEPT AND ITS CLASSIFICATIONS}
 Several concepts of reconfigurable transistors has been reported, as they have been implemented with the low dimensional materials i.e., intrinsic materials, graphene, carbon nanotubes(CNTs)etc \cite{pregl_printable_2016, miryala_verilog-model_2013, nurazzinorizan_carbon_2020} . The concept of RFET depends on either Schottky barrier or the band-to-band tunnelling transistors (TFET). In the case of Schottky barrier, at the metal-semiconductor interface, the transport was constrained by a tunnelling process, also along the source, drain and gate comprised of Schottky junctions, which could directly induce the bending of bands within the Schottky interface, as they are equipped along the carrier injections within the valence and the conduction bands \cite{sessi_junction_2018,zhao_nonvolatile_2021} . On the grounds that, for a bulk semiconductor material, low dimensional geometric materials have been preferred, in the act of promoting the efficiency of electrostatic control over the active regions \cite{gupta_electrostatic_2017} . In the process of achieving a sharp interface on the silicon nanowires between the Al-Si(metal-semiconductor) system  has been facilitated. Hybridization of devices has been reported stating TFET and Schottky barrier field effect transistor (SBFET), as this concept encompasses the negative differential resistance (NDR) as well as impact ionization mode. To facilitate the equal number to electrons and holes, a doping free channel has been implemented in combination with the mid-gap metal electrode in the source and drain, also termed as polarity control, that it could enhanced in such a way that it aids one carrier type depending on the target to be impelled \cite{park_analysis_2022}. Since, all the device concepts are commonly consists of two independent control gate electrodes, whereas the cardinal one is the polarity gate also known polarity gate, which controls the transport mode, at the same time, the other known as the control gate, which switches on/off the transistor.

 \subsection{SCHOTTKY BARRIER}
  Usage of MOSFET shrinks because of the short channel effects and to overcome the short channel effects Schottky barrier tunnel transistor(SBTT) has been preferred for its formation of the ultra-shallow  junction to supress the short channel effects \cite{reuter_mosfets_2020,roemer_physics-based_2022} . Experimental studies alongside with the I/V characteristics reveals that the leakage of current in the SBTT has been concealed using the bulk silicon material, which shows the possibility of producing devices at commercial level \cite{oehme_ge_2010} . SBTT with the thin insulating wafers have been studied along with the p-type as well as n-type, although p-type SBTT has high current performance alongside with the MOSFET \cite{wann_comparative_1996}
  , when compared with the n-type SBTT \cite{kunc_planar_2014} . In the case of SBTT device, the impurity doped source and drain regions are purely replaced by the regions of silicide, which could possibly act as a metal leading to the easy formation of the ultra-shallow junction \cite{baldauf_tuning_2017} . Although the Schottky barrier shows different physical characteristics in comparison with the conventional MOSFET, the flow of carrier particles takes place through the Schottky barrier tunnelling at the silicide interface from the source to the conduction channel \cite{reuter_mosfets_2020, ratnesh_advancement_2021, sakurai_simple_1991, ortiz-conde_review_2002} .
  
In the view of RFET, various RFET principles and architectures along with the Schottky barrier contacts of the intrinsic channels has been reported. Generally, these RFET architecture could be easily make a distinction based on the positioning of the program gates and the bias conditions to be applied \cite{mikolajick_rfetreconfigurable_2017} . Further, another mechanism involved in this mode of differentiating is that by classifying it based on the transport mechanism formed by the control gates, accordingly, this gate drives to a). control over the Schottky barrier as well as the carrier injection and b). complete control over the transport for recently transported injected carriers over a thermionic barrier.

\begin{figure}
    \centering
    \includegraphics[width=10cm,height=9cm]{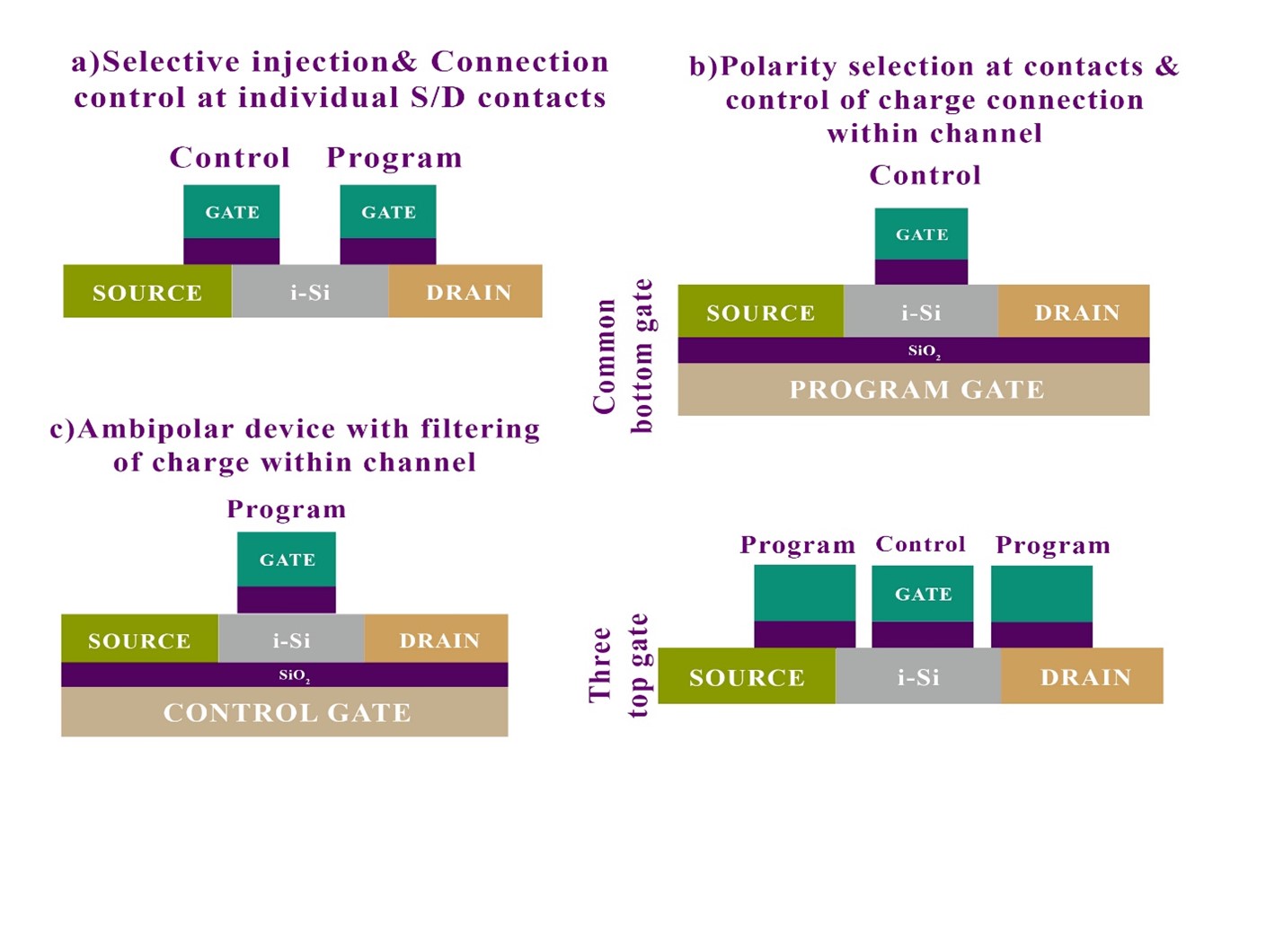}
    \caption{schematic diagram of different concepts of reconfigurable transistors, each one is capable of reconfiguring p and n type characteristics. a). denotes the selective injection control from individual contacts, b). denotes the polarity control at the contacts and the control of charge connection within the channel, c). denotes the ambipolar device with filtering of charge within channel.}
    \label{fig:5}
\end{figure}

As this Schottky barrier RFET’s are commonly built over an intrinsic semiconductor, metal heterostructures and metals, also their basic principle is either a single one or a group of mechanism involved, these mechanism can be functioned efficiently in a single device, further these devices could likely add up all the functionalities via the band to band tunnelling and also the impact ionization \cite{singh_novel_2015} . On the grounds that, in the simpler device instead of using a polarity gate, a common back gate is preferred for reaching the entire channel overlying the Schottky barriers.

The polarity gate is being bend by the energy bands, which thoroughly depends on the voltage being applied. Later the back gate envelopes the complete structure, an electrostatic doping process takes place, as by applying the positive voltage allows the injected electrons to move from the contact regions to the entire channel, likewise, by applying the negative voltage, which allows the injected holes to migrate to the whole channel from the contacts \cite{yu_analysis_2022, gupta_electrostatic_2017, nakaharai_electrostatically_2015} . An additional energy barrier has been regulated by the top control gates in the midpoint of the entire channel, for switching the transistor on and off. In order to streamline the structure as well as to grant the better gating control over the entire device, by using the potentially thinner oxides, the back gate has to be recovered by two additional top gates atop of the Schottky barrier as shown in figure 4. \cite{wang_schottky_2021} . Keeping the applied voltage fixed, the additional two top gates are intended to be the polarity gates synchronically, the control over the type of injected carriers alongside the channel conductance can been altered using the control gate being placed on top of the Schottky barrier \cite{kim_high-performance_2018} .
In the case of an elementary electrostatic doping, the considerable advantage over this doping process is that the control over channel region are not compared by the gates, also it could conceivably operate at low voltages which perhaps reduces the implementation of device fabrication, since all the gates are classically arranged from the front[8]. Further, when  the drain power voltage ($V_D$) is considerably increased beyond the bandgap value of the channel, which drives to steep subthreshold swing at room temperature \cite{zhang_polarity-controllable_2014} .
The injection of the carriers over the Schottky barrier strongly influences the carrier conduction over the material, also the field effect mobility($\mu$) over the material could be accurately hinged with the

\begin{equation}
\mu  = \left (\frac{l}{w} \right ) \left ( \frac{\mathrm{d} }{\epsilon_0\epsilon_r} \right )\left ( \frac{1}{V_D} \right )\left ( \frac{\mathrm{d}I_D}{\mathrm{d} V_D} \right )
\end{equation}

whereas, d denotes the thickness of the gate material, $V_D$ stands for voltage being applied at drain, $I_D$  stands for the drain current, $\epsilon_0$ as the vacuum permittivity, $\epsilon_R$  mainly denotes the dielectric constant \cite{nakaharai_electrostatically_2015} .

In the case of 2D transistor, reconfigurability confides on the Schottky barrier injection, as the study enlightens the idea for optimizing the device along with the structural and material properties of the reconfigurability for the 2D transistors \cite{kang_2d_2020,zhao_nonvolatile_2021} . The on-current levels of the p and n type for a SB-FET strongly relies on the Schottky barrier heights, as to maintain a constant current level as needed to ensure the reconfigurable FET, it should probably have a mid-gap fermi level assembled on the metal-semiconductor contact, considering the current injection is nagged by the Schottky barrier tunnelling, also the on-currents relies on the SB heights. In order to achieve the reconfigurable SB-FETs, large band gap materials such as Si nanowires has been employed as channel materials, accordingly, with the oxides holding the best bandgap values, the threshold voltage has been cut down, which inhibits the shrinkage of operation voltage of the device. In the face of traditional CMOS \cite{navarro_reconfigurable_2017} , these issues could be rectified by using gate materials with multiple work functions as required for the p and n type FETs, when compared the CMOS technique can never been employed to the reconfigurable FETs, as p-FET  and n-FET  are swapped accordingly, also they do have same structure as well as the gate material, therefore the only alternative to lower the threshold voltage value is by picking up the lower bandgap material as channel material, as in consideration of the source injection, but for drain injection, the lower band gap materials should not be employed \cite{xue_overview_2018} .
In the view of 2-D materials, the top and the bottom gate shares the control gate on the source side, while along the drain side an extra bottom gate has been implemented, collectively, these gates improvises the control over the injecting particles \cite{zhang_novel_2022} . In order to combine two or more concepts into a single device, three independent gates along with RFET has been employed, as one gate is programmed to operate on the drain side alone, and the remaining are used to on/off of the transistor. In ambipolar schottky barrier transistors, the degradation of current on/off ratio has been eventually reported, it becomes inevitable for the reconfigurable transistors as it required the mid gap fermi line-up, alongside shifting the fermi level to the conduction band as well as the valence band for n and p FET is ineffective to provide a required solution \cite{reuter_mosfets_2020, ratnesh_advancement_2021, nevoral_ambipolarity_2018, torricelli_ambipolar_2016} . So, as to resolve the issue, a device has been opted to accomplish high on/off current ratio for the reconfigurable field effect transistors with the small band gap materials\cite{nevoral_ambipolarity_2018} . 

Combining two or more mechanism in a single device taken advantage of power saving techniques in the circuitry level. Gate population could be increased exponentially in so as much as the resistance of the Schottky barrier is effective compared with the resistance of the channel materials \cite{trommer_reconfigurable_2016,li_impact_2021} .

\subsection{NON SCHOTTKY BARRIER}
As in the case of reconfigurable FET devices being introduced, which explored the band-to-band tunnelling mechanism as the superior mechanism to achieve reconfigurability, in which the devices comprise of heavily doped silicon regions detached by the intrinsic channel regions. With the sufficient voltage required, the two contact regions can be impelled, which induces the band tunnelling as it allows the electrons to tunnel through the valence band to the conduction band of the intrinsic regions \cite{mikolajick_rfetreconfigurable_2017} . To induce the injection of the carrier particles due to band to band tunnelling a double gate structure has been implemented, concurrently suppressing the ambipolar behaviour of the device, which synchronously leads either of the gates to switch on/off of the device, as the other gate edges to the blockage of polarity charge carriers which has been injected tends to reduce the ambipolar behaviour of the device. In order to achieve reconfigurable FET, the polarity of the device has been switched with the assistance of reversing the gate voltages. Schottky barrier between the injected carrier particles and the channel regions are forbidden, as the conduction does not rely on the thermally generated carriers, also an inferior subthreshold swing has been predicted. The logically switching elements of the device performs very slowly, now that the device suffered from the low on  current  values \cite{trommer_functionality-enhanced_2015, sun_contact_2020} . Considering the contact regions, which are heavily doped, are harder to control on the nanoscale level, also these heavily doped contact regions are essential in order to maintain the band-to-band tunnelling. As in the case of Schottky barrier RFET, the junctions differentiate the tunnel RFET as they employ two gates, although non- Schottky approaches bring about the unipolar characteristics i.e., n or p type depending on the gates intended \cite{zhang_novel_2022, bhattacharjee_impact_2017} . A hybrid concept has been introduced along with the ancillary steering gates near the source and drain contacts could induce electrostatic doping along with the so present tunnelling FET and thermionic switching elements as it enables to operate the device accordingly for tunnel FET as well as thermionic switching \cite{bhattacharjee_first_2017} .

\section{NEGATIVE DIFFERENTIAL RESISTANCE}
The concept of reconfiguration initiated with the concern of reducing the power consumption in the electronics, as negative differential resistance in a circuit could possibly generate a.c. power other than consuming it. Negative differential resistance (NDR) possibly have non-linear electrical properties, these properties are common in insulators with the study of dielectric breakdown \cite{kundu_design_2022, krinke_exploring_2021, jayachandran_reconfigurable_2020} . Zener specified that the breakdown caused by the virtue of quantum mechanical tunnelling within valence and the conduction bands, later he determined that the NDR acts as a switch, the dielectric breakdown acquired from the avalanche breakdown rather than tunnelling \cite{yao_novel_2020} . Further, pure Zener tunnel diode has been fabricated and widely used in the process of obtaining the NDR.

Later on, experiments have been demonstrated with the tunnel diode as it shows a fall in the mobility of the electrons with increasing fields. Specifically, NDR in semiconductors could possibly occurs during various conditions and they are listed below.

\subsection{NDR IN SEMICONDUCTORS}
\begin{itemize}
\item scattering induced NDR, occurs when, there exists more than one scattering processes, as a NDR in parabolic band structure could be formed with the polar longitudinal optic as well as acoustic phonon scattering \cite{ridley_effect_1986} . An instability in the energy balance is caused for the polar longitudinal optic above the critical field and a NDR could be possibly produced by the shift between polar longitudinal optic as well as acoustic phonon scattering.
\item Electron induced NDR, as this confides on the transmission of the electrons from a less denser area to a heavily dense area, NDR arises with the rapid rise in the density of states, which tends to produce an non-parabolic NDR \cite{gaudioso_vibrationally_2000} .
\item Optically induced NDR, optical excitation of electrons takes place with the well-defined levels in the conduction band, using the monochromatic light \cite{ridley_negative_1993} . On the grounds that, the rate of capture is very less compared to that of the scattering induced negative differential resistance.
\item Magnetically induced NDR, in a magnetic field, all electrons lies within the landau  level, therefore transport of electrons takes place via the spatially separated states through scattering processes \cite{singh_thermal_2019}. Negative mobility effect has been neglected and the probabilities of filled states of the initial and final states is taken into account \
\cite{wisniewski_anomalous_2022} .
\item Field enhanced NDR, the mechanism involved in this type of inducing the NDR is that an impurity barrier effect is associated for the capture of impurities around the potential barrier. In field enhanced NDR, the field effect mobility changes otherwise there will be a reduction in the field effect mobility owing to the effect \cite{sharma_graphene_2015} .
\item Double injection induced NDR, this method of inducing is designed in such a way that the current in the system is being determined by the injection of electrons in the cathode as well as injection of holes in the anode. A current controlled NDR arises as if the recombination centre is completely filled with the electrons. As in this case, injected electrons are not easily trapped when compared with the injected holes. Comparatively, at high currents, the concentration of the injected holes equals the injected electrons as it leads to the smooth capturing of holes rather than the electrons.
\item The major cause of NDR in double injection induced NDR is the transition takes place between the two effective carrier injections, also NDR could possibly occur when either of the carrier injection is blocked, as this induces tunnelling process within the carrier injection \cite{guo_switches_1998} . The I/V characteristics clearly represents the current controlled NDR while it shifts from the single carrier curve to two carrier curves.
\item Transit- time induced NDR, as this mechanism involves operating at zero frequency, as there exists several numbers of device along with the impact ionization diode that has been operated using the NDR with microwave frequencies. Diodes of high field regions, where the injected carriers generate avalanche along with the impact ionization \cite{smith_grating-induced_2015} .
\end{itemize}

\subsection{NDR INSTABILITIES}
On the grounds that, a material exhibiting NDR is electrically unstable, simultaneously, voltage controlled NDR leads to the formation of high field domain i.e., either propagating or stationary \cite{hall_negative_1984} . Current spike, being generated by the propagating domains, these spikes vanish at the anode and the period of oscillation is observed by the transit time, which can be determined by trapping. A fascinating phenomenon occurs, when the domain velocity got reduced by trapping leads to attain the active acoustic wave velocity i.e., piezoelectric along with the direction of propagation, as this case induces large acoustic strains. 
  
1. Acoustoelectric instability, occurs in the form of damped or continuous, oscillations at frequencies in respect to the transit time along with the transversely polarised acoustic modes, also the drift velocity at the threshold field for the instabilities are greater than the transverse acoustic modes.

2. High field oscillations, generally occurs at room temperature in the sample materials containing 3D electron densities. The threshold field over the high field oscillations strongly depends on the carrier concentration as well as the barrier height and the well width, but the frequency of the oscillation is independent of the length and field of the samples.

3.Destructive instabilities, as in the case of samples containing large electron densities, the instabilities are caused by the current oscillations which could cause irreversible destruction of the sample.

4.Carrier depletion, undoped materials with the lattice temperature below 10K  current collapses simultaneously carrier depletion occurs at the electric fields, consequently followed by the current oscillation frequency \cite{da_cunha_ndr_1993}.

\section{ENHANCING REFT’s TECHNOLOGY}
To establish RFET in a highly sophisticated level, distinct basic design and material key elements are considered independent of concept chosen also precluding the working principle of the system \cite{mikolajick_rfetreconfigurable_2017,da_cunha_ndr_1993,sun_contact_2020}. One of the distinct features of RFET is the existence of two or more independent gates, as this specific feature profoundly characterises the reconfigurability over the devices, though an additional gate is required for switching between the operational modes. When considering traditional MOSFET \cite{sakurai_simple_1991}, the major stumbling block is that higher area is opted for MOSFET to produce the same amount as individual transistors. Turning the drawbacks literally to benefits has been undertaken as an challenge over this feature in terms of functionality, for instance, reconfigurability over the devices could be achieved with the assistance of basic circuitry functions like NAND and NOR, also XOR along with some minority gates constructed with lesser number of transistors \cite{trommer_inherent_2020} . In the view of three gated structures, polarity control has been treated as mode control as differences in threshold voltage has been studied along with power saving techniques cognate by three independent gate devices \cite{saha_nanowire_2019} . Coexisting, sub-threshold slopes at 6 mV/dec has been demonstrated in sequence of impact ionization along with the positive feedback effect, offers an excellent option for analog designing. The comparison over the device performance is not sufficient for upgrading the RFET technology, also the independent gates are not just the design feature but also corelates with properties of the materials. The inception of the RFET devices intended over the 1-dimensional nanostructures like germanium nanowires, carbon nanotubes (CNT’s) and also with silicon nanowires, as these materials possess high surface quality, the major add-on advantage over the material is identical distribution over narrow structures, being the better contender for device demonstration [51,86]. The advancements over RFET technology led to factory-oriented manufacturing enduring silicon on insulator (SOI)-based RFET in collaboration with classic complementary metal oxide field effect transistor (CMOS-FET), etched nanowires. As of the performance, back end of line integration seems a better option. 
For RFET’s along with NDR properties, Germanium has been the most astonishing channel material, as it encloses the potential for embellishing power efficiency as well as the speed of RFET’s. For a top-down fabrication, Ge based RFET has been utilized as Ge on insulator wafers, also GeSi based RFET utilizes SiO2 on the insulator wafers \cite{quijada_germanium_2022, roemer_physics-based_2022} .
Materials exhibiting electrostatic control at high current density theoretically in specific, 2D layered systems like graphene, black phosphorous and dichalcogenides, functioning as band structures depends on the thickness of the materials, bond free surfaces, stacking of various insulating materials also conducting and semiconducting property over the device enables optimizing properties over the material \cite{chen_overview_2020,benedek_symmetric_2023,  voiry_phase_2015} . 
On the grounds of organic RFET’s which has been discussed over, precluding the material used, RFET’s requires more trivializing channels with good gate control to aid with the high current density and very less applied voltage. Also, the applications over the organic RFET’s are located in the regions of flexible substrates \cite{krinke_exploring_2021, wang_flexible_2021} .

\begin{figure}
    \centering
    \includegraphics[width=9cm,height=7cm]{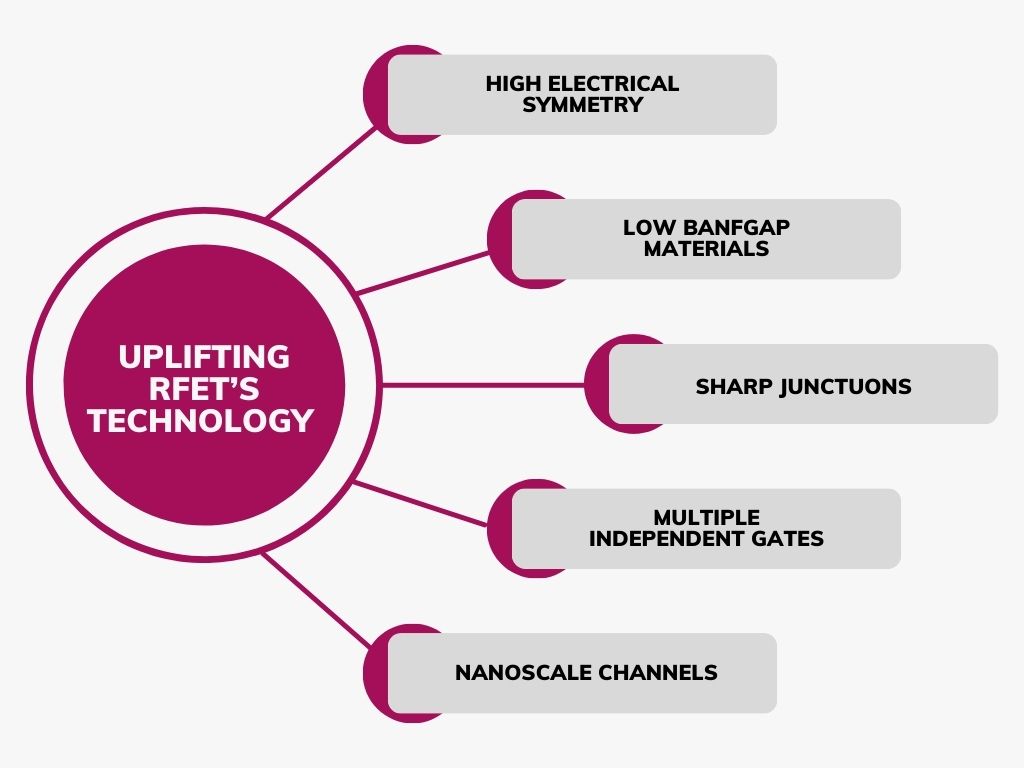}
    \caption{visual examples for enhancing the RFET’s technology for the successful RFET fabrication}
    \label{fig:6}
\end{figure}

In the motive of enhancing device to another level, gate control throughout the device has been uplifted with the reinforcement of sharp junctions, as in the case of Schottky barrier-based devices, sharp junctions has been utilized by the metal silicide or germanides \cite{kundu_design_2022, tang_solid-state_2014, weber_silicon-nanowire_2006} . The process involving silicidation can even produce metal/semiconductor interface ensuring atomic level sharpness, also $NiSi_2$ has been reported for silicon devices, correspondingly it influences epitaxial affinity with silicon as mentioned in figure 5. considering the monolithic Al-Ge-Al heterostructures \cite{bockle_gate-tunable_2021, rajan_novel_2022,bockle_gate-tunable_2021, mikolajick_reconfigurable_2022} fabricated through top-down synthesis contributes stable crystal phase also assuring a predictable, yet decisive reproducible contacts. Enduring the contacts imposed in the semiconductor nanowires or nanosheets, carrier injection with high precision could be restrained with the aid of gate contacts at the junction \cite{quijada_germanium_2022, duan_dispersive_2021, mikolajick_silicon_2013} . Ethnologically, the major confronts along the control over the position of silicide or germanide formation for the device consistency. Excluding the position of the contact regions, electrical properties over the contact surfaces must be supervised and restrained, as majority of the metal semiconductor combinations shows a variety of phases also has to be controlled. An alternate solution is required to achieve sharp junctions, in the case of no formation of metalide contacts over the channel materials \cite{gupta_electrostatic_2017} .

In the consideration of 2D materials, if metalide doesn’t form a sharp junction, it could be rectified by designating the source and drain electrodes on the top of the channel as the gates are laid down, this could induce a thin back gate dielectric formation which could be associated with the carrier injection at very low operational voltages. In addition, the integration over RFET’s along with classic CMOS requires constant current over the conduction modes as in electrons and holes \cite{alasad_logic_2017, weber_reconfigurable_2014, knechtel_hardware_2020} . The major drawback on using the CMOS is that individual optimization over both the device types prosecuted in classic CMOS is not possible. Consequently, p-type and n-type enhances the device level technologically, if it doesn’t withhold proportionate amount, there will never be an upliftment on the device level, instead there is shift in the switching point away from the actual one leads to sagging of the multiple logic states above on another has been disrupted. Also, this could be tenaciously resoluted by the help of high scaled electronics, which is operated at very low operational voltages. Experiments have been carried out in the prospective of attaining symmetry over the Si- based RFET devices. An obligatory need of attention is intended for the constant current flow throughout the source, drain and gate. Materials like $NiSi_2$/ Si, an effort has been made to attain symmetry over the source and drain contacts, in the case of silicon-based polarity controllable devices \cite{khan_controlled_2021,mikolajick_silicon_2013,sistani_nanometer-scale_2021} . In order to find a secondary effect used to regulate the symmetry over the electron and hole currents, persuading self-limiting oxidation say $Si_3Ni_4$ materials. Measures have been taken for those materials that doesn’t rely on the formation of thermally alloyed contacts, for instance, materials with different work functions at the source and drain have been contemplated for current matching. An alternate approach over this is that utilizing different work function material within the common gate which is likely to alter the configuration of the blocking potential present within the channel.

Enhancing the RFET technology, symmetric n and p type operation plays an eminent role, as well as specific optimization over the transport of both the carrier types are not viable, whereas the channel material should opt for both operational modes. In the usage of low bandgap materials like Ge, InAs reportedly showcases excellent performance notably on lower threshold voltage \cite{cadareanu_parasitic_2021} . So, in order to utilize comparatively low bandgap materials, carrier injection takes place via tunnelling mechanism, also simultaneously increasing the electric field over the channel material. To the same degree, devices based on Schottky barrier, the effective performance can be achieved by the combined barrier height for electrons and holes adding up to the bandgap value \cite{xue_overview_2018} . Off-state leakage currents, massive increase with decreasing bandgap. Comparatively, silicon-based devices allow lowest off-state leakage current values because of its wide bandgap than the germanium-based devices. As in this case, reconfigurable FET mitigates the drawback over the low bandgap materials by blocking the carrier type, also carrier type has been altered to generate the RFET \cite{reuter_mosfets_2020,zhang_novel_2022} . Ge based devices shows prominent off state in Al-Ge contacts than the counterparts with $Ni_2$ Ge-Ge, as this could be attributed to the strong fermi level in the Al-Ge system near the valence band. Considering graphene materials, regardless of the absence of bandgap, could be passed down to realize the polarity control devices employing double gate structure \cite{nevoral_ambipolarity_2018} , also layered transition metal dichalcogenide even could yield bandgap lower than silicon.

\begin{table}[ht]
\caption{Comparison of electrical metrics for different reconfigurable device concepts}
\label{tab:table1}
\resizebox{\columnwidth}{!}{%
\begin{tabular}{@{}llllllll@{}}
\toprule
 \textbf{\shortstack{ Channel material \\ \& Orientation}} & \textbf{Contacts} & \textbf{\shortstack{Channel length \\ (nm)}} & \textbf{\shortstack{On Current \\ p-program}} & \textbf{\shortstack{On Current \\ n-program}} & \textbf{\shortstack{Off current of \\ n \& p program}} & \textbf{\shortstack{Subthreshold slope \\ n/p (mV/dec)}} & \textbf{Reference} \\ \bottomrule  
 Si (100) & NiSi & 100-200 & 7 $\mu$A & 4 $\mu$A & 1 pA & 64/70 & Zhang et.al  \cite{zhang_schottky-barrier_2014} \\
    Si (112) & NiSi$_2$ & 680 & 1.9 $\mu$A & 110 nA & 3 fA & 90/220 & Singh et.al \cite{singh_novel_2015} \\
    Si (110) & NiSi$_2$ & 220 & 135 $\mu$A & 135 nA & 50 fA & 150/150 & Weber et.al \cite{weber_silicon-nanowire_2006} \\
    Si & NiSi & 1000 & 100 $\mu$A & 50 nA & 1 pA & 343/428 & Mongillo et.al \cite{mongillo_multifunctional_2012} \\
    Graphene & Ti/Au & N/A & 30 $\mu$A & 70 pA & 1 pA & 583/1538 & Sharma et.al \cite{sharma_graphene_2015} \\
    CNT & Ti & 900 & 2 nA & 100 pA & 1 pA & 63-134/173 & Norizan et.al \cite{nurazzinorizan_carbon_2020} \\
    Si (110) & NiSi$_2$ & 480 & 20 pA & 10 pA & 0.1 pA & 375/240 & Zhang et.al \cite{zhang_polarity-controllable_2014} \\
    Si & NiSi & 6000 & 0.8 $\mu$A & 1.4 $\mu$A & 0.4pA & 100/150 & Heinzig et.al \cite{park_high-density_2017} \\
    Si & SiO$_2$ & N/A & 15.2 $\mu$A & 128 $\mu$A & N/A & 187 & Chen et.al \cite{chen_overview_2020} \\
    Ge & Al & 150 & 13.6 $\mu$A & 12 $\mu$A & N/A & 64 & Cadareanu et.al \cite{cadareanu_parasitic_2021} \\
    Si & Pd/Ni & N/A & 5 nA & 5 nA & 4 pA & 130 & Roy et.al \cite{roy_dual-gated_2015} \\ \bottomrule
\end{tabular}%
}
\end{table}

\section{ADD ON FEATURES}
Ethically, the trend elevating over years focusing on the betterment of the device on its circuitry level, in addition increasing the versatility of the RFET devices. In the majority, the widely discussed concept is the merger of polarity control factor along with the non-volatile storage choice \cite{zhao_nonvolatile_2021} . In consideration of neuromorphic circuitry applications alongside with the transparent and flexible computing, the need to develop the merging between the polarity control and non-volatile storage becomes an inevitable one as the functions over the neuromorphic computing in memory designing requires highly desired functions to enable the enduring perpetual program over the circuit's functionality \cite{nakaharai_electrostatically_2015, park_high-density_2017} . 
Multi-valued logic (MVL) \cite{zhang_schottky-barrier_2014, gonzalez_standard_2000} alongside with the RFET devices has been utilized as an preference for increasing the functionality by replacing the traditional binary systems with the operational pattern holding greater base. The prejudice over the statement is that MVL are in NDR regions alone. Reports suggested that the NDR regions in the Al-Ge-Al nano structure, showcases prominent features of polarity control of RFET, as this encourages the potential of creating circuit designs inclusive of RFET’s \cite{baldauf_tuning_2017, park_reconfigurable_2018} .

\section{POLARITY CONTROL WITH NON-VOLATILE MEMORY}

RFET concept has been originally developed to regulate them reversibly over the polarity of transistors over the run-time. The polarity control over the device could be easily obtained by the application of supplementary static voltage \cite{cappelletti_non_2015,ouyang_programmable_2004} , which has been maintained consistently. The supplementary static voltage, called as static voltage selects the carrier type which passes through the Schottky barrier i.e., either electrons or holes \cite{gupta_electrostatic_2017,singh_novel_2015} . Although, when the voltages are no longer applied, the information is lost and the operation turns volatile \cite{han_towards_2013, ouyang_programmable_2004, poremba_nvmain_2015} . Since the polarity gate requires a clear-cut description but embedding the non-volatile memory with RFET just not only bring add on additional advantages in terms of multi-operational memory functions as well as close concurrency within the logic and memory enabling new computing archetype \cite{lankhorst_low-cost_2005,bae_high-density_2017} . To integrate the functionality over the operating system by adding up the memory preserved layers like charge trapping as well as ferroelectric materials within the gate load of FET. 
 
As per the literature reports, plenty of experimental studies has been carried out in the field of non-volatile RFET’s. The foremost work was carried out by schwalke et al., they use charge trapping layer, particularly buried oxide of SOI sample by applying high voltage to the back gate, as this could not be suitable for single devices while programming \cite{park_high-density_2017, chatterjee_methodology_2009} . Though, charge trapping requires high applied voltages for switching between p and n type, but till date it is reported consisting capable of smaller memory \cite{lee_demonstration_2022}. 

Followed by, park et al., \cite{park_high-density_2017} uses programmable back gate and oxide $/$ nitride $/$ oxide stack to introduce non-volatile memory systems, as this offers plenty of operational modes but requires complicated processing \cite{han_towards_2013, lankhorst_low-cost_2005, chatterjee_methodology_2009, heinzig_dually_2013,weber_reconfigurable_2014,  lee_demonstration_2022}. Advancing it to the other level, park et al., bottom down grown Si nanowire was utilized along with the metal nitride oxide (NMO)at the gate stack and it reports successful programming and erasing of p and n-type modes, as they could enable 4 different operation in a single device \cite{frohman-bentchkowsky_metal-nitride-oxide-silicon_1970,frohmanbentchkowsky_charge_2003}.
 
Deliberately, advancing the features, electrostatic doping has been preferred by ferroelectric polarization has been briefed below. Accordingly, experiments have been carried out to introduce the non-volatility to the silicon nanowires SB-FET by a thin doped Hafnium Oxide (HfO$_2$) film into the gate stack. Experiments revealed that junction transmissibility can be tuned with the help of program gate \cite{ju_high_2016,lee_nonvolatile_2020}.

The considerable drawback are the limited reproducibility and scalability of the devices fabricated using the bottom-up method to grown silicon nanowires \cite{gonzalez_standard_2000}. As this could be rectified later on by replacing bottom-up approach with the top- down approach to synthesize silicon nanowires and it has been successfully unified with the Hafnium Zirconium Oxide(HZO) along with the stack of single nanowire SB-FET \cite{kunc_planar_2014}. The on currents over the ferroelectric polarization has been tuned and this empowered a better memory for the hole conductance.

\section{NDR mode RFET}

\begin{figure}
    \centering
    \includegraphics[width=15cm,height=9cm]{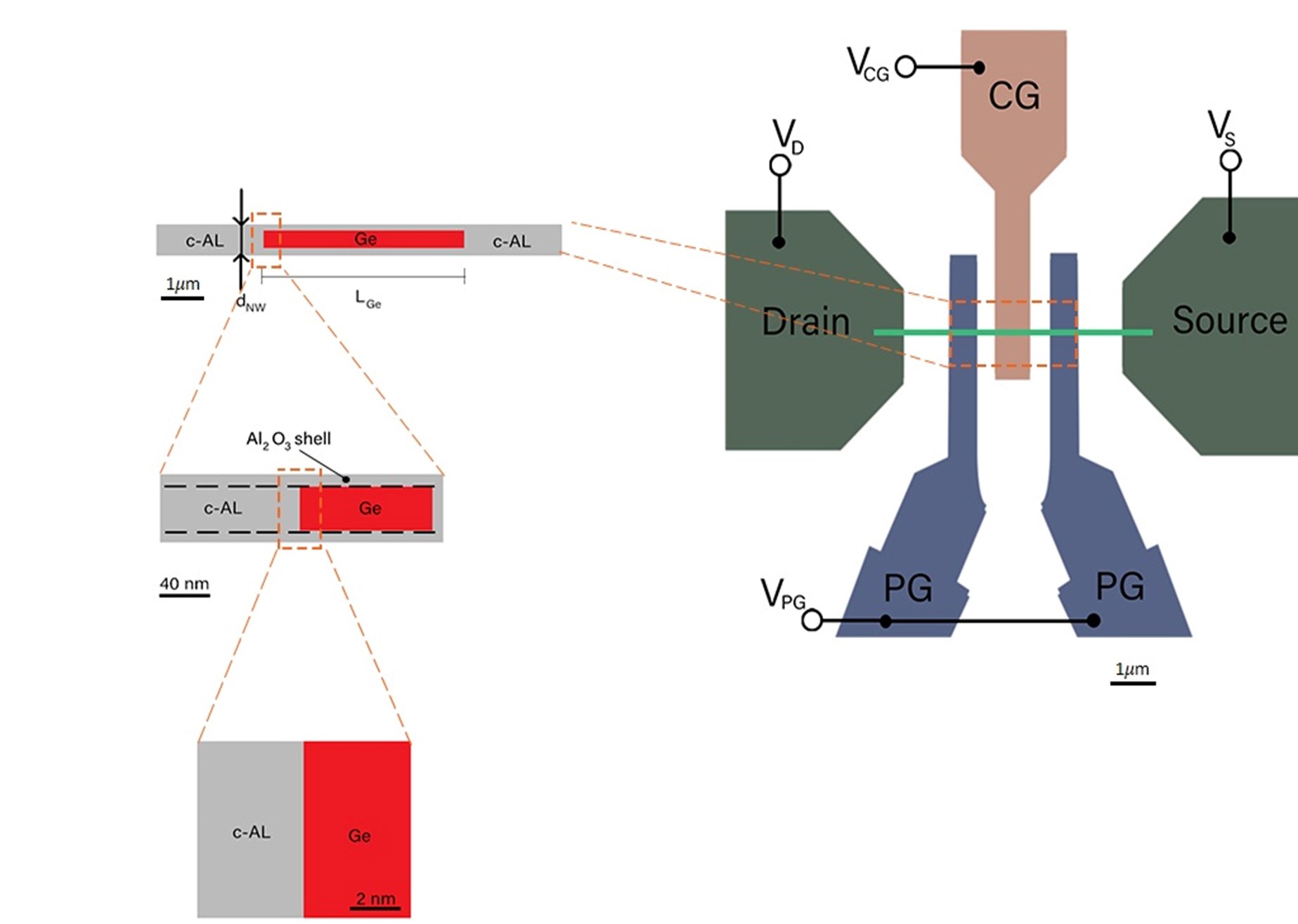}
    \caption{a. represents the schematic sketch of HAADF STEM image of the Al-Ge-Al nanowire heterostructure, b. represents the schematic sketch of HR-TEM images showcasing the whole Al-Ge interface of the nanowire heterostructure, c. represents the schematic diagram of the SEM image for the Al-Ge-Al heterostructure withholding the three gates, which could enable RFET operation along with the NDR functionalities.}
    \label{fig:7}
\end{figure}

Presentably, NDR mode RFET concept competent of exhibiting both the p and n type operations inclusive of gate tunable NDR functionality as regulated by programmed applied voltages, which accounts in enhancing the primitive switching units, permissive to the computational as well as analog oscillatory circuits. Taken into consideration quasi 1D monolithic Al-Ge-Al nanowire heterostructure exhibiting comparatively low operational voltages along with the reconfigurable polarity overhead i.e., p and n type operations, along with the approachability over the transferred electron effect at the room temperature as mandatory for providing the features of NDR mode over RFET devices \cite{sistani_room-temperature_2017} . So, as to facilitate p and n type reconfigurable FET as well as gate tunable functionalities, three independent top gates has been assigned with the Al-Ge-Al nanowire heterostructures. With the utilization of the device architecture, Al-Ge-Al heterostructure has been accomplished to incorporate the electrical properties over the unipolar n and p type FETs within an individual device inclusive of identical technology, geometry as well as composition, the complexity over the device technology could be reduced gradually, as they could enable dynamic and runtime programming.  Also, the Al-Ge junction serves as a polarity gate in setting the charge carriers to switch between p and n type operation, simultaneously, the control gate could be utilized to modulate the thermionic drain current through the transistor, also allowing the device to turn on or off. Setting VPG = -5V, p type operation has been programmed, as the transfer measurements were demonstrated as such VCG between -5 to +5 v for the fixed bias voltages between VD = -100mV and 1V followingly, the p type operation results in the upward bending beneath the PG’s which could enable the hole injection into the Ge channel \cite{gan_voltage-controlled_2010} .

Governing the thermionic emission of the injected charge carriers through the Ge channel via the VCG. High variation over the different applied drain voltages VCG has been observed over the on-currents of the n type operation. Likewise, on applying VPG =5V, fixing the junction with some higher transparency of electrons, leading to downward bending resulting in the enhancement of electron injection due to the barrier becoming thinner. The high Schottky barrier over the Al-Ge heterostructure contacts cannot be achieved in other typical metal contact to Ge \cite{tung_recent_2001}.

On the view of L- and $\Gamma$- minima are firmly close to energy, depending on the calculation of the coupling constants in between the conduction band valley in bulk Ge. Experimental works reveal that the L- and X-minima rather than L- and $\Gamma$- minima with certain effective masses. Linearly increasing the VD alongside monitoring the ID for various V$_CG$ of 0 to 5 V could eventually affirms the voltage controlled NDR features with the advancing enhancement for higher VCG \cite{ortiz-conde_review_2002, gan_voltage-controlled_2010}.  In order to characterize the peak-to-valley ratio (PVR), defined by I(Vpeak)$/$I(Vvalley), also a stable PVR could highly relate to many logic applications. Reconfigurability over the NDR mode with its temperature dependence by applying the control gate voltage V$_CG$= 5V, also with its hiking thermal excitation and considerable electron concentration over the L- valley, the peaks present in the NDR region has been transferred from the lower electric fields to the higher ones. The PVR values decreases with increase in temperature when there occurs any decrease in the transfer efficiency caused by the least scattering mean free path which could be correlated to the thermal excitation of charge carriers.

 Also, in comparison with the Ge quantum dot tunnelling diodes along with the Si and Si/SiGe resonant inter-band tunnelling diodes, at room temperature PVR of our best RFET devices is comparatively a factor 10 larger. Besides, the most desired device behaviour could be adequately used for small foot-print and energy efficient computational multivalued logic (MVL) \cite{guo_switches_1998, cadareanu_parasitic_2021,frohman-bentchkowsky_metal-nitride-oxide-silicon_1970} , as the radix is set by the number of devices connected in series. Most precisely, the device concept enables a modulation of both the PVR and the position of the NDR peak, also along with the cascade of devices, individually biased with a different voltage on the control gate, would aftereffect   results in I$/$V characteristics comprising protruding NDR regions consisting of MOBILE devices, enabling both NAND and NOR operations \cite{trommer_functionality-enhanced_2015,heinzig_dually_2013, mongillo_multifunctional_2012, wei_ge-based_2018, bae_reconfigurable_2019} . Alongside, NDR mode RFET concept could advance enabling applications such as static memory cells, fast switching logic circuitry and energy efficient computational MVL.
\section{RFET POTENTIAL AND ITS FUTURE APPLICATIONS}

Reconfigurable FET’s, along with its significance of flexibility and transparency has gained much attention as it acts as a bright candidate for numerous future applications like neuro-inspired computing as well as hardware securities, also with the current applications such as light sensors, primitive based biosensors and charge-based biosensors \cite{saha_nanowire_2019} . RFET concept over neuro-inspired computing as well as hardware security has been briefed. Comparably, RFET’s along with the added systems might not be powerful enough to replace the conventional CMOS entirely \cite{japa_hardware_2021} . RFET, with its advantageous applications can be used as an add on functionality, but when co-integrated along with the classic CMOS, system level performance could be enhanced. Also, by virtue of biosensors, RFET concepts has been extensively studied, as the biosensing applications include health, agriculture and food industries etc. the devices are capable enough to detect various biological stimulus and convert them into electrical signals, further RFET devices along with the biosensing applications had gained much attention because of its compact size, rapid response timing, reduced sample volume and assimilation of the circuits processed \cite{ghasemi_advances_2023, han_single_2021} . Beyond the bounds, the reconfigurability holds lots of possibilities for the mixed signals as well as analog signals. Also an interesting concept to be covered, is the cryogenic electronics for quantum computing as it overcomes the traditional quantum computing, also cryo CMOS devices could operate at deep cryogenic and excellent performance at low temperatures which facilitates error correcting loops, readout and control and integration with qubits \cite{aoki_performance_1987} . The merging of reconfigurability and the NDR provides an excellent platform for variety of applications which includes  fast switching multi valued logic, photo detectors with efficient dark current suppression, high frequency oscillators \cite{jayachandran_reconfigurable_2020, heinzig_dually_2013} .

\subsection{ADVANCES WITH NEURO-INSPIRED COMPUTING}

An emerging field of research, brain inspired neuromorphic computing with the diverse advancements related to the artificial intelligence has energy efficient computing than the classical von-Neumann structure \cite{mariantoni_implementing_2011} . The leverage in using the neuromorphic computing is, it can perform intelligent tasks with relatively low power consumption and with reduced chip size. Cognitive processes such as recognition of patterns, analysis of speech and the prediction of system’s behaviour can be carried out efficiently using the artificial neural networks (ANN) \cite{pavan_analysing_2023} , and are surpassing than the classic computing hardware, which mimic the cognitive process. Complementary metal oxide semiconductors (CMOS) are being replaced for the purpose of neuromorphic computing on the grounds that it requires plenty of transistors to implement the synaptic as well as the functionality of neurons because of its excessive consumption of power and of high cost \cite{schwalke_cmos_2013} . Except the biological counterpart, all states are less energy efficient, some of the properties of biological system with the standard systems includes a). Merging of storage and computing, b). adaptability to the environment, c). well organized and compactible. Advancing non-volatile memory devices have paved attention to the researchers for the implementation of the neuromorphic computing, devices such as Ferroelectric field effect transistor (FeFET) has attracted for its high energy efficiency, high performance and diverse synaptic behaviour like human brain. Neuromorphic computing \cite{wang_flexible_2021, yu_neuro-inspired_2018, furber_large-scale_2016} , any which way changes the function by the electrical stimulation, RFET provides the capability to adopt the changes and their functionalities, also paved a way in the development of plastic biologicals components like neurons and synapses at the device level.

\begin{figure}
    \centering
    \includegraphics[width=11cm,height=9cm]{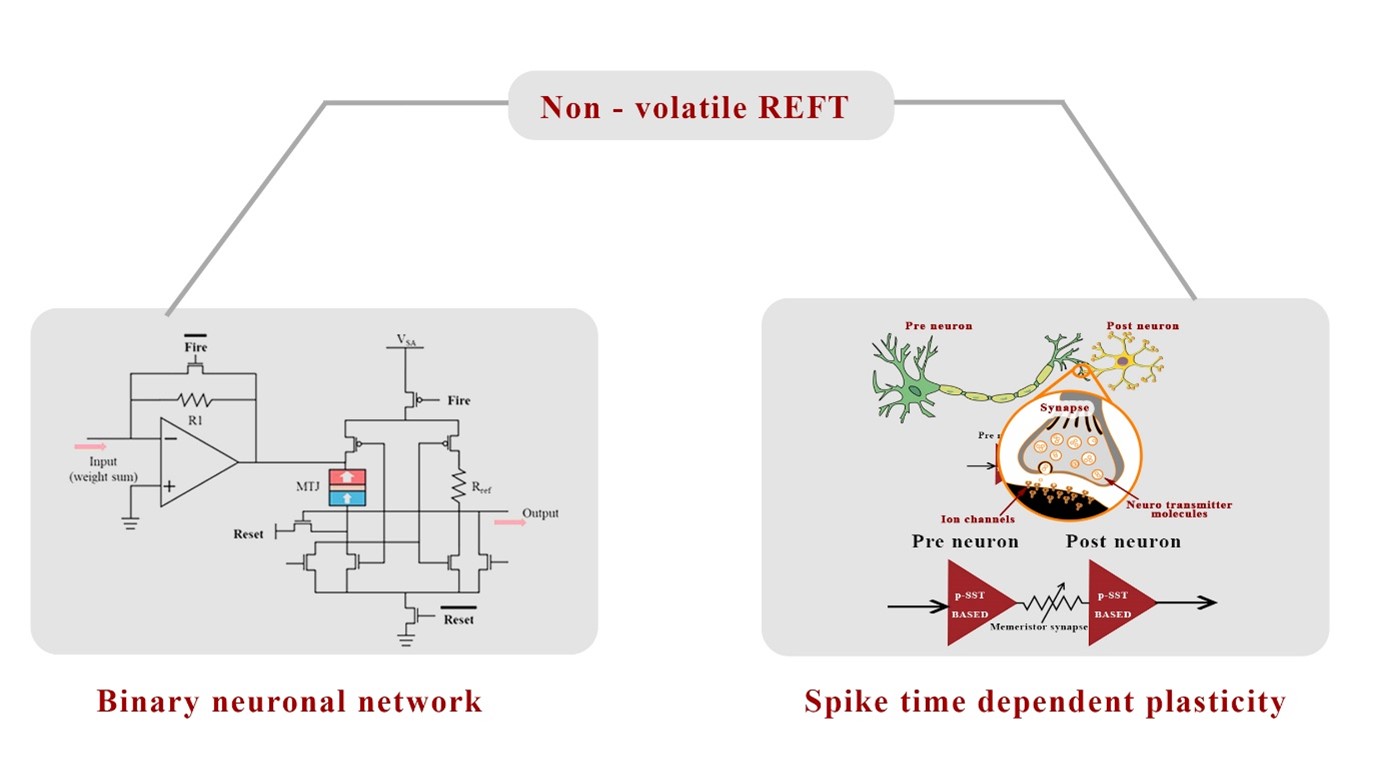}
    \caption{represents the neuro-inspired computing. a). circuit diagram of simple binary neuronal network. b).diagram of synapse capable of realizing  the spike time dependent plasticity.}
    \label{fig:8}
\end{figure}

To enhance the field effect characteristics of the reconfigurability logic or neuromorphic computing, diverse complex electronic circuits have been opted for the reconfiguration, instead by developing the grain level reconfigurability unit at device as simplified circuitry level would enhance the field effect characteristics.

Fengben et.al., stated that the ferroelectric partial polarization has been employed for the artificial synapses, so that the switching functionalities progressively make modifications in the $NiSi_2$/Si Schottky barrier. Synaptic functions such as paired pulse facilitation and short-term synaptic performance have been exhibited by the as produced synaptic devices, also the amplitude of the pre synaptic spike has been corelated with the amplitude of the excitatory post-synaptic current \cite{xi_artificial_2021} .

Danijela et.al., stated that the synapses as well as neurons performance can be enhanced by implying the oxide materials in the electronic devices, the Von Neuman concept stands forward as an issue in the artificial neural networks, which includes performing data with complicated process and bringing back the data to the memory \cite{markovic_analysis_2001} .
Tian-Yu Wang, et.al., notes that the long-term plasticity plays a vital role in the memorizing, learning and recognition and the mechanical flexibility of the device must be inspected, as it leads to wearable neuro inspired computing application. This article showcases the $HfAlO_x$ based device which shows high density binary storage, multibit resistive switching and excellent conductance required for the neuromorphic computing \cite{wang_flexible_2021} .

 Mitigating power consumption over the neuromorphic devices and storage of charges are mandatory for the efficiency and performance of the devices, yet some of the sparing performance is required, such as large network accessibility for read and write, mass production at low cost, auto assembling capability. To obtain the biological proportionate, various preparation techniques has been adopted and reconfigurability does not simply rely on particular components, instead numerous combinations has been implemented to satisfy the reconfigurability of the devices \cite{van_de_burgt_organic_2018} .

\subsection{HARDWARE SECURITY}

As the advancements in the device manufacturing leads to integrated circuit production along with the astounding semiconductor intellectual property(IP), which gravitates to avoid piracy over hardware’s IP \cite{chen_using_2016, bi_emerging_2016} . In the view of this issue, a technique known as hardware obfuscation has been implemented, on the point that it protects the IP inherent in the product, as the other companies tends to product’s reverse engineering, which allows the competitors to understand the product. To the same degree, it could happen in industries ranging from computer software, automotive industry as well as electronics \cite{japa_hardware_2021} . 

 In order to obtain the product infornmation, the specialist handles different sort of techniques like a). visual inspection of integrated circuits(IC’s) which includes exposing the IC and obtaining a set of images through scanning electron microscope(SEM) across the layers of it, followed by a process called electrochemical polishing (EMP),essentially it detects the information of IC’s just like by which components it is made and how they are connected to other, b).side channel attacks,  these are nothing but completely monitoring the behavior of the device i.e., power consumption and time taken for the execution of the task and c).controlled fault injection, causes faults in the system’s operation, as it tends to reveal the internal details of the system, followed by altering the power supply over IC’s , clock frequency and also it can even change the operating temperature of the system \cite{knechtel_hardware_2020, chen_using_2016} .
 
 \begin{figure}
    \centering
    \includegraphics[width=16.5cm,height=7cm]{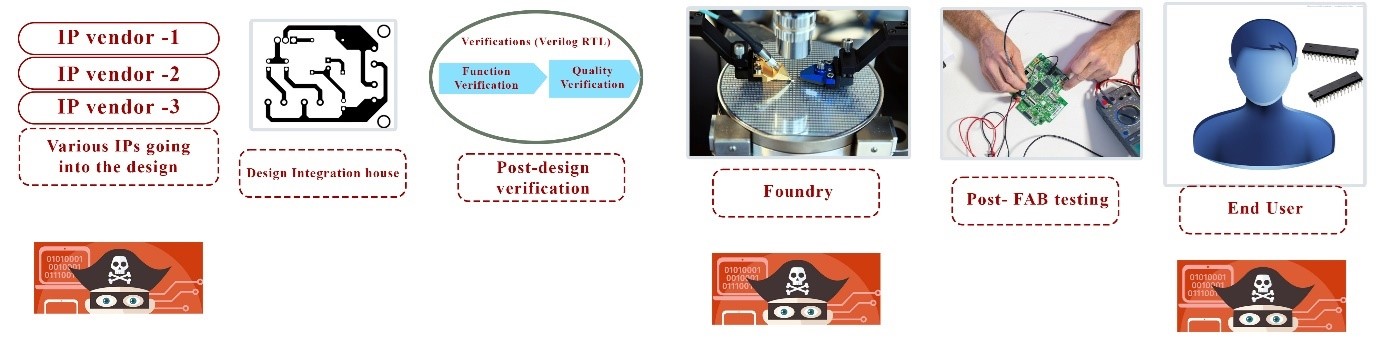}
    \caption{security for the global IC supply chain, as listed various attack measures has been used at mainly IP vendors, foundry and end user}
    \label{fig:9}
\end{figure}

 Since there is an increasing demand for protecting the hardware security from the infringement, tending to strengthen the hardware level security. Consistently, numerous techniques has been opted for protecting the IP has been listed down as split manufacturing, watermarking, camouflaging and logic locking pattern, etc. \cite{yasin_hardware_2019} 
Specifically, logic locking pattern uplifts the hardware security across the channel of the IC supply chain, anyhow it is an establishment between the higher security and the area, and it cost atop. Since, there is a boundless focus on the development over the new algorithms for security measures as well as logic locking \cite{sengupta_atpg-based_2018} . Various imminent devices have been considered as a promising candidate for replacing the traditional CMOS commensurate logic gates.
Amongst the emerging devices, RFET acts as an eminent candidate, which compliments the traditional CMOS gates because of its predominant characteristics listed as, functional polymorphism and structural polymorphism, alongside polymorphic performance facilitates new appeals to the hardware security solutions quick-fix as physical unclonable functions, chip authentication, etc. RFET based on logic cells and flip-flops are less spaced out to delay the side channel attacks compared with the CMOS and its counterpart, revealing the potential for hardening the system over the outside attacks\cite{markovic_analysis_2001} . Also, an additional advantage of using RFET is that REFT’s over runtime configuration can enable building the polymorphic logic gates, contradictory to the CMOS equivalent logic gates, as RFET could perform more than one function. As in the case of RFET’s logic locking, the functionality of the polymorphic gates can be reconfigured depending on the applied key values and the logic locking based on RFET has been discussed \cite{mehta_hardware_2016} .
RFET’s based polymorphic gates are prominent enough to replace the logic gates of the original netlist, which possibly have locking key which could enable the reconfigurable functionality over the circuit. Aforesaid, functional locking has added advantage over the traditional CMOS based logic locking. In the case of post fabrication reconfigurability, it is hard to reverse engineer the layout and obtain the original netlist for other companies because the uniform and planar layout of the as-produced fabricated design remains camouflaged \cite{patnaik_advancing_2018} . As the RFET based polymorphic gates minimizes the area overhead and power-delay product substantially as compared to the CMOS based logic locking design.

Above discussed, RFET’s along with its unique behavior from the traditional CMOS, as well as a combination of both to form hybrid IC’s would add an additional level of variables, which tends to make the attacks on intellectual property much difficult yet expensive than the usual \cite{bi_enhancing_2016} . 

A simple SOI along with the RFET could be readily inserted into the traditional CMOS ICs without any cost, as they could elevate to next level for the reverse engineering process, as they tends to be highly expensive. Alongside, an interesting evolution emerged from the polymorphic logic gates with the three independent gate reconfigurable field effect transistor (TIG-RFET’s), having a potential of using it based on multiplexer for the logic locking \cite{baldauf_stress-dependent_2015} . As this could also navigates to the increase in key length as per the logic locking gate by using multi gated transistor architecture.

Apart from logic locking, RFET also shows various potential over the hardware security. Similar to the CMOS gates, RFET’s has the possibility of split manufacturing. The exact functionality of the polymorphic gates with RFET could be examined during the post fabrication process.

Another hardware security measure is the watermarking, as the RFET based inverters has been preferred as an embedding watermarking scheme within the circuit. Due to reconfigurability, the RFET based inverter performs invariable logic as the inverter maintains the functionality of both values of the program gate voltages. So, this enables the designer to use any kind of encoding scheme and can be embedded into the inverter with the very strong watermark \cite{zhang_novel_2022} . Further with the dynamic reconfigurability, latches based on the RFET’s can maintain two metastable stages within a clock cycle, as this tends to generate double the random numbers over a single clock cycle, when compared with the CMOS based latch. From the works that has been carried out enlisted that the RFET‘s can be exclusively utilized for the new advancements of hardware security to prevent them against the attack schemes.

\section{CONCLUSION}

Over a long period of time, RFET concepts have been introduced and developed by shrinking the device at its dimension, in order to meet the demand over high speed, high compatibility and low operating power, as NDR mode RFET with its huge advantages makes it easier to meet the demand of devices being precise and compact, which could be multifunctional. Early stage of the works was wrapped with the demonstrations and later on, extended alongside with the device optimisation and circuitry applications. An ideology on how to fabricate optimized devices as well as implementation of circuits at advance level with significant advancements over the traditional CMOS. An entire changeover is suggested for the complete manufacturing of RFET devices as it requires consequential alterations in the design methodology as well as flow. Lately, applications based on the integration of RFET’s along with the traditional CMOS are of significant concern, as this could enable an industrial application over the device materials and such devices has been successfully fabricated with very low process integrated along with the CMOS process. It is imperative that RFET’s can be used as an add on functionality rather being substituted for classical devices, as an application, hardware security as well as neuro-inspired computing can be a trend setting attractive applications throughout. Alongside, the applications extend to two dimensional devices adding on features like non-volatility and yet circuitry applications approaching the run time configuration. As NDR mode RFET further extend its complexity over numerous applications with integrated circuits with the help of novel research approaches and yet lot more applications to be at its leading edge in the upcoming years.

\section*{ACKNOWLEDGEMENT}

I would like to express my profound gratitude to Manonmaiam Sundaranar University for being a constant underprop in the accomplishment of my article.

\bibliographystyle{ieeetr}
\bibliography{ref}

\end{document}